\pdfoutput=1

\documentclass[11pt]{article}

\usepackage[final]{acl}

\usepackage{times}
\usepackage{latexsym}

\usepackage[T1]{fontenc}

\usepackage[utf8]{inputenc}

\usepackage{microtype}

\usepackage{inconsolata}

\usepackage{graphicx}

\usepackage{amsmath}
\usepackage{amsthm}
\usepackage{booktabs}
\usepackage{algorithm}
\usepackage{algorithmic}
\usepackage[switch]{lineno}
\usepackage{booktabs}
\usepackage{fontawesome}
\usepackage{tikz}
\usepackage{inconsolata}
\usepackage{pifont}
\usepackage[edges]{forest}
\usepackage{makecell}
\usepackage{cleveref}

\usepackage{enumitem}
\usepackage{setspace}
\usepackage[utf8]{inputenc}
\usepackage{url}
\usepackage{booktabs}
\usepackage{soul}
\usepackage{contour}
\usepackage{tcolorbox} 
\usepackage{marvosym}

\usepackage{graphicx}
\usepackage{tikz}
\usepackage{forest}
\usetikzlibrary{trees,positioning,shapes,shadows,arrows.meta}
\usepackage{footmisc}

\title{Mastering the Craft of Data Synthesis for CodeLLMs}

\author{
    Meng Chen\textsuperscript{\Letter} \enskip Philip Arthur \enskip Qianyu Feng \enskip Cong Duy Vu Hoang \\
    \textbf{Yu-Heng Hong \enskip Mahdi Kazemi Moghaddam \enskip Omid Nezami \enskip Thien Nguyen} \\
    \textbf{Gioacchino Tangari \enskip Duy Vu \enskip Thanh Vu \enskip Mark Johnson} \\
    \textbf{Krishnaram Kenthapadi \enskip Don Dharmasiri \enskip Long Duong \enskip Yuan-Fang Li\textsuperscript{\Letter}} \\
    \textsuperscript{\Letter} \texttt{\{meng.c.chen, yuanfang.li\}@oracle.com} \\
    Oracle Corporation
}

\begin{document}
\maketitle

\begin{abstract}
 
Large language models (LLMs) have shown impressive performance in \emph{code} understanding and generation, making coding tasks a key focus for researchers due to their practical applications and value as a testbed for LLM evaluation.
Data synthesis and filtering techniques have been widely adopted and shown to be highly effective in this context. In this paper, we present a focused survey and taxonomy of these techniques, emphasizing recent advancements. We highlight key challenges, explore future research directions, and offer practical guidance for new researchers entering the field.
\end{abstract}


\section{Introduction}\label{sec:intro}

Code intelligence leverages machine learning techniques to enhance software development by improving both code quality and programmer productivity \cite{allamanisminingsoftware, allamanissurvey}. The rise of LLMs, such as ChatGPT ~\cite{gpt35}, Gemini \cite{geminiteam2024geminifamilyhighlycapable}, Claude \cite{claude}, and Llama \cite{dubey2024llama3herdmodels}, has significantly reshaped the automation of code-related tasks, including code completion \cite{guo2023longcoderlongrangepretrainedlanguage}, translation \cite{szafraniec2023code}, repair \cite{olausson2024is}, and documentation \cite{khan2022automaticcodedocumentationgeneration}. Tools like GitHub Copilot \cite{chen2021evaluatinglargelanguagemodels}, CodeGeeX \cite{zheng2023codegeex}, and Cursor \cite{cursor} hold great promise in substantially increasing human programmer efficiency and revolutionizing the software industry, attracting considerable attention from both academia and industry. Recently, specialized LLMs for code-related tasks (denoted as CodeLLMs) have emerged, including Code Llama~\cite{roziere2024codellamaopenfoundation}, StarCoder~\cite{li2023starcodersourceyou,lozhkov2024starcoder2stackv2}, DeepSeek-Coder~\cite{guo2024deepseekcoderlargelanguagemodel, deepseekai2024deepseekcoderv2breakingbarrierclosedsource}, and CodeQwen~\cite{bai2023qwen}.

Recent advancements \cite{gunasekar2023textbooks, gandhi-etal-2024-better} in LLMs have highlighted the critical role of high-quality data in building strong, robust models. Similarly, for CodeLLMs, diverse, high-quality datasets are essential for improving performance across a wide range of code-related tasks. Significant efforts have been devoted to collecting and curating code-related corpora. Prominent examples include the Pile~\cite{gao2020pile}, the Stack~\cite{kocetkov2023the,lozhkov2024starcoder2stackv2} and BigScience ROOTS~\cite{laurenccon2022bigscience}, which draw primarily from open-source and permissively licensed platforms such as GitHub and Stack Overflow.

However, relying solely on human-generated data for code-related tasks poses several challenges. First, collecting large-scale human data is labor-intensive and expensive, particularly for high-quality instruction tuning and preference alignment data. Second, human-generated data is prone to biases and errors \cite{hosking2024human, singh2024beyond}, as it reflects the varying skill levels of programmers, and may not be optimal for model training. Third, data integrity concerns, such as the risk of sensitive personal/corporate information leakage, complicate data collection. Lastly, for low-resource programming languages—--either due to limited popularity or proprietary restrictions—--data scarcity hinders the effectiveness of CodeLLMs in specialized fields and systems programming \cite{mora2024syntheticprogrammingelicitationrepair}. Consequently, synthetic data generated by LLMs has emerged as a valuable alternative to complement natural data. Leveraging their vast knowledge and advanced linguistic capabilities, LLMs can generate high-quality data, providing a valuable foundation for model training in code-related tasks.

\begin{figure*}[ht]
    \centering
    \setlength{\abovecaptionskip}{2mm}
    \setlength{\belowcaptionskip}{-5mm}
    \includegraphics[width=0.73\textwidth]{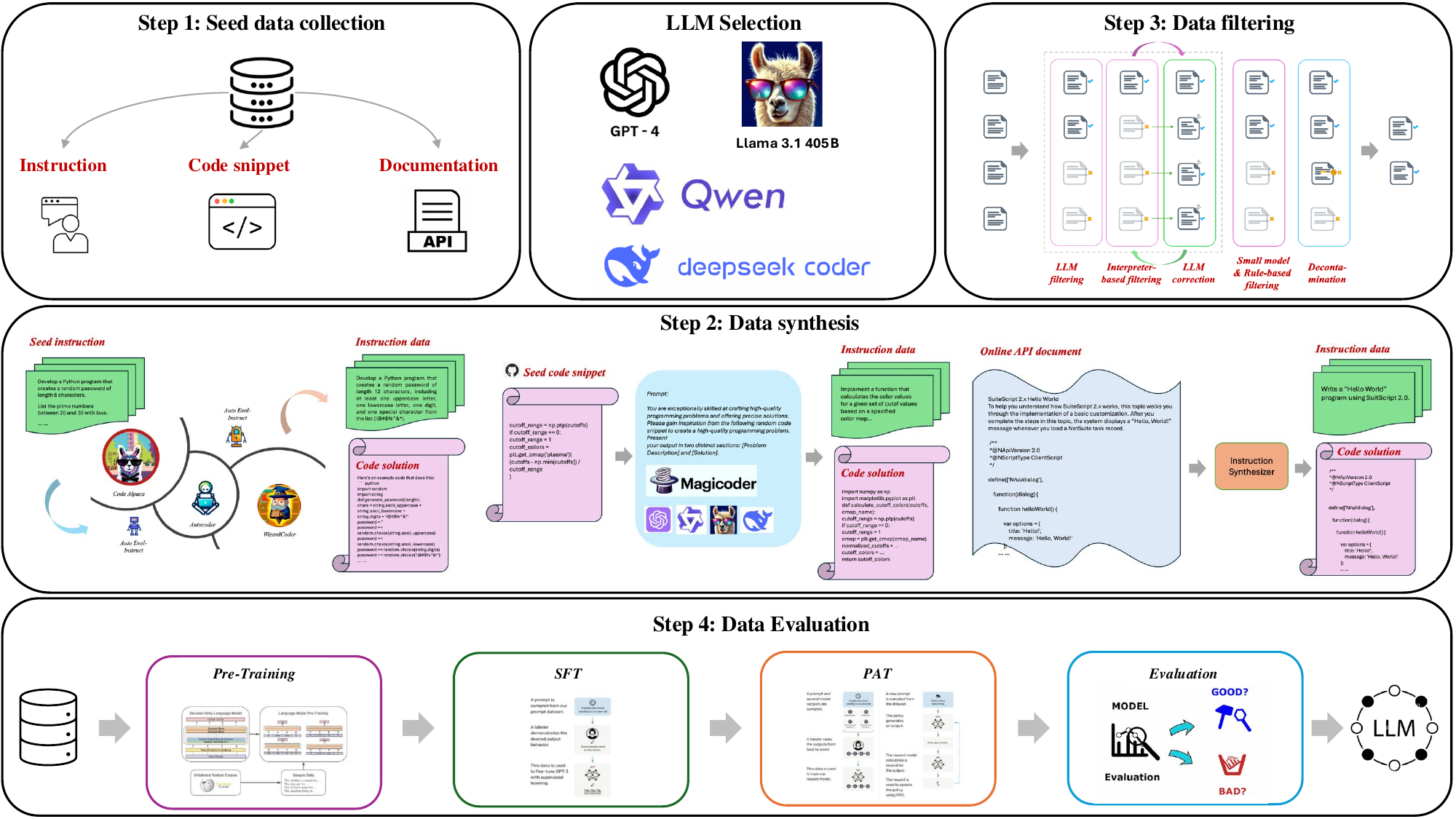}
    \caption{Practical guidance for the code related data generation pipeline. We also recommend several large language models (LLMs) that offer strong performance while maintaining a balanced cost (Appendix~\ref{sec:llm_section}).}
    \label{fig:guidance}
\end{figure*}

While generating synthetic datasets for code-related tasks may appear straightforward, achieving both high accuracy and sufficient diversity is a complex process requiring meticulous design and advanced techniques \cite{gandhi-etal-2024-better}. This makes a systematic exploration of LLM-driven synthetic data generation both essential and timely. Although there are survey papers in the fields of general data engineering \cite{liu2024bestpracticeslessonslearned, long-etal-2024-llms, wang2024datamanagementtraininglarge, ding-etal-2024-data} and code intelligence \cite{wan2023deeplearningcodeintelligence, jiang2024surveylargelanguagemodels, zhang2024unifyingperspectivesnlpsoftware, sun2024surveyneuralcodeintelligence}, there is a notable gap in literature focusing specifically on data synthesis and filtering techniques for code-related tasks. To fill this gap, we present a targeted review of recent advancements in synthetic data generation and filtering for training CodeLLMs, covering over 50 recent works across 23 topic categories from the past two years. The techniques discussed are organized into a taxonomy (Fig.~\ref{fig:lit_surv}) and analyzed in terms of their motivation, methodologies, and key contributions. 
We also maintain a GitHub repository\footnote{\label{repo}\url{https://github.com/chenmengdx/awesome-data-synthesis-for-code-llm}} to collect open-source synthesis datasets for CodeLLMs and track recent advancements.
Our goal is to provide an in-depth overview of the current state of the field, highlight key challenges, and offer insights to guide researchers and practitioners in building efficient and robust CodeLLMs through effective data engineering practices.

\section{Preliminaries and Related Works}\label{sec:overview}


\subsection{The Data Curation Pipeline}\label{sec:pipeline}
Data curation, which aims to ensure datasets are of high quality, diverse, relevant, and available, is crucial to the success of CodeLLMs. The data curation process typically involves four key steps (cf.\ Figure~\ref{fig:guidance} in ~\ref{sec:guidance}). \textbf{(1) Seed Input Collection}: Before synthesizing data, a small set of seed samples (e.g.\ problem-solution pairs), unlabeled inputs (e.g.\ code snippets), or human-written instructions (e.g.\ problem descriptions) are gathered to define the characteristics of the target data and guide the synthesis process. \textbf{(2) Data Synthesis}: LLMs are leveraged to generate a large volume of code-related data samples for specific downstream tasks, exploiting their comprehensive coding-related knowledge and capabilities. \textbf{(3) Data Filtering}: This step involves removing low-quality, irrelevant or redundant samples, addressing issues such as hallucinations or ambiguous descriptions caused by ineffective prompts, to ensure the dataset's usefulness. \textbf{(4) Data Evaluation}: The final step assesses the quality and applicability of the data to confirm its value for downstream tasks. 

\vspace{-4pt}
\subsection{Relationship to Other Works}

\noindent\textbf{Data Synthesis \& Selection.} Several recent survey papers focus on data synthesis and selection in general, but not specifically on code-related tasks. \citet{liu2024bestpracticeslessonslearned} track the state of synthetic data research, outlining best practices and key lessons learned. \citet{long-etal-2024-llms} address the lack of a unified framework in LLM-driven synthetic data generation, proposing a general workflow by organizing studies around generation, curation, and evaluation. \citet{wang2024surveydataselectionllm, albalak2024a} provide a thorough review of recent advancements in data selection methods. \citet{xu2024surveyknowledgedistillationlarge} present a comprehensive review of knowledge distillation, structured around algorithms, skills, and verticalization, and explore distillation mechanisms, cognitive skill enhancements, and their practical applications across various domains. \citet{wang2024datamanagementtraininglarge} offer an extensive overview of data management strategies in both pretraining and supervised fine-tuning stages of LLMs. \citet{ding-etal-2024-data} analyze the impact of LLMs on data augmentation, while \citet{tan2024largelanguagemodelsdata} review learning strategies for models using LLM-generated annotations. Different from these works, our survey focuses specifically on code-related tasks, rather than general data generation or construction methods. 

\noindent\textbf{Code Intelligence.} Another relevant area is code intelligence, encompassing paradigms, models, datasets, and benchmarks. \citet{she2023pitfallslanguagemodelscode, zan-etal-2023-large, wan2023deeplearningcodeintelligence, jiang2024surveylargelanguagemodels, zhang2024unifyingperspectivesnlpsoftware, sun2024surveyneuralcodeintelligence, zhang2024surveylargelanguagemodels, lyu2024automaticprogramminglargelanguage} provide general reviews of advances in code intelligence, particularly in code generation. \citet{liu2024surveynl2sqllargelanguage} present a comprehensive analysis of LLM-based NL2SQL techniques, covering the entire lifecycle---model, data, evaluation, and error analysis. \citet{zhang2024systematicliteraturereviewlarge} conduct a systematic literature review of LLM applications in automated program repair. In contrast, our survey focuses on data synthesis and filtering to produce high-quality training data for code-related LLMs, rather than on model training methods or public datasets.

\begin{figure*}
    \centering
\tikzset{
    basic/.style = {draw, rounded corners=2pt, text width=2cm, align=center, font=\sffamily\tiny, rectangle},
    root/.style  = {basic, thin, align=center, fill=white, minimum height=2cm, text width=0.2cm},
    onode/.style = {basic, thin, align=center, fill=green!30, text width=0.2cm},
    tnode/.style = {basic, thin, align=left, fill=pink!60, text width=2cm, align=center},
    xnode/.style = {basic, thin, align=left, fill=cyan!20,text width=6cm},
    wnode/.style = {basic, thin, align=left, fill=pink!10!blue!80!red!10, text width=6em},
    edge from parent/.style={draw=black, edge from parent fork right},
}

\begin{forest} 
forked edges,
for tree={
    grow=east,
    growth parent anchor=west,
    parent anchor=east,
    child anchor=west,
    edge path={\noexpand\path[\forestoption{edge}] 
         (!u.parent anchor) -- +(10pt,0pt) |- (.child anchor)
         \forestoption{edge label};},
    anchor=center, 
    font=\sffamily\tiny,
    ver/.style={basic, rotate=90, child anchor=north, parent anchor=south, anchor=center},
}
[Data Synthesis and Filtering for Code-related Tasks, ver, l sep=5mm, text width=5cm
    [Data Filtering, ver, fill=green!30, l sep=5mm,
        [Decontamination, tnode,  l sep=7mm,
            [Semantic-level matching, wnode,
                [{Abstract syntax tree \cite{riddell2024quantifyingcontaminationevaluatingcode}, Embedding similarity \cite{ding2024semcodertrainingcodelanguage}}, xnode,]
            ]
            [Surface-level matching, wnode, 
                [{Substring matching, Hashing matching \cite{li2023starcodersourceyou,lozhkov2024starcoder2stackv2}}, xnode]
            ]
        ]
        [LLM-based Filtering, tnode,  l sep=7mm,
            [LLM-as-a-Judge, wnode,
                [{ALPAGASUS \cite{chen2024alpagasustrainingbetteralpaca}, ICE-Score \cite{zhuo2024icescoreinstructinglargelanguage}, LLM discriminator \cite{yu-etal-2024-wavecoder}, Model-as-judge \cite{dubey2024llama3herdmodels}}, xnode]
            ]
        ]
        [Small Model-based Filtering, tnode,  l sep=7mm,
            [Model-based, wnode,
                [{Coreset selection \cite{chen2023maybe05dataneeded, yu-etal-2024-wavecoder}, Classifier-based \cite{dubey2024llama3herdmodels}}, xnode]
            ]
            [Indicator-based, wnode,
                [{Difficulty score \cite{li2024superfilteringweaktostrongdatafiltering}, Instruction quality \cite{cao2024instructionmininginstructiondata}, CodeBERTScore \cite{zhou2023codebertscoreevaluatingcodegeneration}}, xnode]
            ]
        ]
        [Interpreter-based Filtering, tnode,  l sep=7mm,
            [Execution-based, wnode,
                [{SemCoder \cite{ding2024semcodertrainingcodelanguage}, AutoCoder \cite{lei2024autocoderenhancingcodelarge}, SC2-Instruct \cite{jiaweiliu2024starcoder2instruct}}, xnode]
            ]
            [Parser-based, wnode,
                [{Dependency Parsing \cite{guo2024deepseekcoderlargelanguagemodel, deepseekai2024deepseekcoderv2breakingbarrierclosedsource}}, xnode]
            ]
        ]
        [Rule-based Filtering, tnode,  l sep=7mm,
            [De-duplication, wnode,
                [{String matching and MinHash \cite{lee2022deduplicatingtrainingdatamakes}, File-level and repository-level \cite{ deepseekai2024deepseekcoderv2breakingbarrierclosedsource}, Global \& local deduplication \cite{shen2024slimpajamadcunderstandingdatacombinations}}, xnode]
            ]
            [Heuristic rules, wnode,
                [{Basic filters (long line filter, alpha filter, encoded data filter) \cite{li2023starcodersourceyou, lozhkov2024starcoder2stackv2}, Language-specific filter \cite{guo2024deepseekcoderlargelanguagemodel, deepseekai2024deepseekcoderv2breakingbarrierclosedsource}}, xnode]
            ]
        ]
    ]
    [Data Synthesis, ver, fill=green!30, l sep=5mm,
        [Specific Tasks, tnode,  l sep=7mm,
            [Code Documentation, wnode,
                [{CodeExp \cite{cui2022codeexpexplanatorycodedocument}, DistillCodeSum \cite{su2024distilledgptsourcecode}}, xnode]
            ]
            [Code Factoring, wnode,
                [{Performance-Improving Edits \cite{shypula2024learningperformanceimprovingcodeedits}}, xnode]
            ]
            [Code Translation, wnode,
                [{Back Translation \cite{chen2023exploringdataaugmentationcode}}, xnode]
            ]
            [Unit Test Generation, wnode,
                [{Actor-Critic RL \cite{gorinski2023automaticunittestdata}}, xnode]
            ]
            [Code Repair, wnode,
                [{SemCoder \cite{ding2024semcodertrainingcodelanguage}, DebugBench \cite{tian2024debugbenchevaluatingdebuggingcapability}, DistiLRR \cite{chen2024personaliseddistillationempoweringopensourced}}, xnode]
            ]
            [NL2SQL, wnode,
                [{SENSE \cite{yang2024synthesizingtexttosqldataweak}, AmbiQT \cite{bhaskar2023benchmarkingimprovingtexttosqlgeneration}, DR. Spider \cite{chang2023drspiderdiagnosticevaluationbenchmark}, ScienceBenchmark \cite{zhang2023sciencebenchmarkcomplexrealworldbenchmark}}, xnode]
            ]
        ]
        [Core Objectives, tnode,  l sep=7mm,
            [Iterative Programming, wnode,
                [{OpenCodeInterpreter \cite{zheng2024opencodeinterpreterintegratingcodegeneration}, SemCoder \cite{ding2024semcodertrainingcodelanguage}, CYCLE \cite{ding2024cyclelearningselfrefinecode}, LETI \cite{wang2024letilearninggeneratetextual}, Reflexion \cite{shinn2023reflexionlanguageagentsverbal}}, xnode]
            ]
            [Enhance Reasoning, wnode,
                [{LLM-Assisted Code Cleaning \cite{jain2023llmassistedcodecleaningtraining}, SemCoder \cite{ding2024semcodertrainingcodelanguage}, CodePLAN \cite{sun-etal-2024-enhancing-code}, BeyondCode \cite{cao-etal-2024-beyond}, Case2Code \cite{shao2024case2codelearninginductivereasoning}}, xnode]
            ]
            [Strengthen Diversity, wnode,
                [{Magicoder \cite{wei2024magicoderempoweringcodegeneration}, Auto Evol-Instruct \cite{zeng2024automaticinstructionevolvinglarge}, WaveCoder \cite{yu-etal-2024-wavecoder}, LintSeq \cite{piterbarg2024traininglanguagemodelssynthetic}}, xnode]
            ]
            [Improve Quality, wnode,
                [{LLM-Assisted Code Cleaning \cite{jain2023llmassistedcodecleaningtraining}, PERsD \cite{chen2024personaliseddistillationempoweringopensourced}, Self-play \cite{haluptzok2023languagemodelsteachprogram}, AutoCoder \cite{lei2024autocoderenhancingcodelarge}, Llama 3.1 \cite{dubey2024llama3herdmodels}},xnode]
            ]
        ]
        [Building Phases, tnode,  l sep=7mm,
            [Model Evaluation, wnode,
                [{CRUXEval \cite{gu2024cruxevalbenchmarkcodereasoning}, AmbiQT \cite{bhaskar2023benchmarkingimprovingtexttosqlgeneration}, DR. Spider \cite{chang2023drspiderdiagnosticevaluationbenchmark}, ScienceBenchmark \cite{zhang2023sciencebenchmarkcomplexrealworldbenchmark}}, xnode]
            ]
            [Preference Alignment, wnode,
                [{CodeUltraFeedback \cite{weyssow2024codeultrafeedbackllmasajudgedatasetaligning}, PLUM \cite{zhang2024plumpreferencelearningplus}}, xnode]
            ]
            [Supervised Fine-Tuning, wnode,
                [{Code Alpaca \cite{codealpaca}, WizardCoder \cite{luo2023wizardcoderempoweringcodelarge}, Magicoder \cite{wei2024magicoderempoweringcodegeneration}, Auto Evol-Instruct \cite{zeng2024automaticinstructionevolvinglarge}, WaveCoder \cite{yu-etal-2024-wavecoder}, SemCoder \cite{ding2024semcodertrainingcodelanguage}, AutoCoder \cite{lei2024autocoderenhancingcodelarge}, MultiPL-T \cite{cassano2024knowledgetransferhighresourcelowresource}}, xnode]
            ]
            [Pre-training, wnode,
                [{Phi-1 \cite{gunasekar2023textbooks}, Phi-1.5 \cite{li2023textbooksneediiphi15}, CodeLlama \cite{roziere2024codellamaopenfoundation}, Instruct PT \cite{cheng2024instructionpretraininglanguagemodels}}, xnode]
            ]
        ]
    ]
]
\end{forest}
    \caption{Taxonomy of data synthesis and filtering techniques for code-related tasks.}
    \label{fig:lit_surv}
\end{figure*}

\section{Key Data Synthesis Techniques}\label{sec:synthesis}

This section reviews recent data synthesis techniques for code-related tasks, structured by the taxonomy in Figure \ref{fig:lit_surv} along three dimensions: Building Phases, Core Objectives, and Specific Tasks. \emph{Building Phases} categorizes works by stages of CodeLLM construction, including pre-training, fine-tuning, alignment, and evaluation. \emph{Core Objectives} groups studies by goals like enhancing data quality, increasing diversity, improving reasoning, and supporting iterative programming. \emph{Specific Tasks} include NL2SQL, code repair, unit test generation, translation, refactoring, and documentation.

\subsection{Model Building Phases}

\noindent\textbf{Pre-training.} A notable example among code LLMs is the \textit{Phi} series, which is primarily trained on synthetic ``textbook-quality'' data. This includes less than 1B tokens of GPT-3.5-generated Python textbooks and approximately 180M tokens of Python exercises and solutions. The \textit{Phi} models, such as \textit{Phi-1} \cite{gunasekar2023textbooks} for Python coding and \textit{Phi-1.5} \cite{li2023textbooksneediiphi15} for commonsense reasoning and language understanding, outperform many open-weight models on coding benchmarks like HumanEval \cite{chen2021evaluatinglargelanguagemodels} and MBPP \cite{austin2021programsynthesislargelanguage}, despite being 10 times smaller in model size and 100 times smaller in dataset size. This demonstrates the effectiveness of synthetic data in training. CodeLlama \cite{roziere2024codellamaopenfoundation} generates about \~{}14,000 Python question-test-solution triplets by first creating unit tests and then verifying generated solutions. 
\citet{cheng2024instructionpretraininglanguagemodels} propose augmenting corpora with instruction-response pairs generated by an instruction synthesizer, followed by continual pre-training on the augmented data. Trained this way, Llama3-8B outperforms Llama3-70B in some cases.

\noindent\textbf{Supervised fine-tuning.} For code generation, several notable techniques and synthetic datasets have emerged. Code Alpaca \cite{codealpaca} introduces a dataset of 20K code instructions, generated via the SELF-INSTRUCT method \cite{wang-etal-2023-self-instruct} applied to ChatGPT across 21 seed tasks. WizardCoder \cite{luo2023wizardcoderempoweringcodelarge} enhances the complexity of code instructions, using the Evol-Instruct technique \cite{xu2024wizardlm}, resulting in a dataset of 78K evolved code instruction examples. To address inherent biases in LLMs and foster diverse, creative code instructions, Magicoder \cite{wei2024magicoderempoweringcodegeneration} employs ChatGPT to generate 75K diverse synthetic instruction samples inspired by random open-source code snippets. \citet{zeng2024automaticinstructionevolvinglarge} introduces Auto Evol-Instruct, an end-to-end framework that evolves instruction datasets using LLMs without manual intervention. WaveCoder \cite{yu-etal-2024-wavecoder} compiles the CodeSeaXDataset, consisting of 19,915 instruction instances that integrate task definitions and associated requirements, covering tasks such as code summarization, generation, translation, and repair. SemCoder \cite{ding2024semcodertrainingcodelanguage} curates PYX, a collection of 34,639 executable code samples with functional descriptions and execution traces. AutoCoder \cite{lei2024autocoderenhancingcodelarge} introduces AIEV-INSTRUCT, a two-stage agent interaction framework that constructs 169K high-quality code instruction samples. Fine-tuned on this dataset, AutoCoder outperforms GPT-4 Turbo and GPT-4o in pass@1 on the HumanEval benchmark. \citet{cassano2024knowledgetransferhighresourcelowresource} introduce MultiPL-T, an effective approach for generating semi-synthetic data for low-resource programming languages using test-validated translation of high-quality code in high-resource languages.

\noindent\textbf{Preference alignment.} \citet{weyssow2024codeultrafeedbackllmasajudgedatasetaligning} present CodeUltraFeedback, a preference dataset comprising 10,000 complex instructions and 40,000 responses generated by 14 diverse LLMs, aimed at aligning LLMs to coding preferences in code generation scenarios. \citet{zhang2024plumpreferencelearningplus} propose PLUM, a preference learning framework for training CodeLLMs. It uses GPT-4 to generate unit test cases from natural language instructions, samples candidate solutions, and evaluates them against the test cases to create a preference dataset of \~{}180K samples.

\noindent\textbf{Evaluation.} \citet{gu2024cruxevalbenchmarkcodereasoning} develop CRUXEval (Code Reasoning, Understanding, and eXecution Evaluation), a benchmark consisting of 800 Python functions created using a ``generate-and-filter'' approach with CodeLlama. \citet{bhaskar2023benchmarkingimprovingtexttosqlgeneration} introduce AmbiQT, a novel benchmark with over 3,000 examples where each natural-language question can be interpreted as two plausible SQL queries due to lexical and/or structural ambiguity. This benchmark is generated through a combination of ChatGPT-based synonym generation and perturbation, along with standard rule-based perturbation. \citet{chang2023drspiderdiagnosticevaluationbenchmark} curate Dr.Spider, a comprehensive diagnostic robustness evaluation benchmark with 15K perturbed examples generated by paraphrasing natural questions. ScienceBenchmark \citet{zhang2023sciencebenchmarkcomplexrealworldbenchmark} is a complex NL2SQL benchmark for three real-world scenarios, created by extending a small amount of human-generated data with synthetic data using GPT-3.

\subsection{Core Objectives}
\noindent\textbf{Quality.}
Ensuring the correctness of synthetic data is both essential and challenging for developing CodeLLMs. \citet{jain2023llmassistedcodecleaningtraining} introduce a novel pipeline to improve the dataset quality by enhancing code structure and readability. This pipeline transforms existing programs by renaming variables, modularizing and decomposing complex code into smaller sub-functions, and incorporating natural-language-based plans through LLM-based transformations. PERsD \cite{chen2024personaliseddistillationempoweringopensourced} employs a personalized distillation process to improve data quality through adaptive refinement, leveraging the student's generated code and its execution feedback. \citet{haluptzok2023languagemodelsteachprogram} propose enhancing CodeLLMs using a self-play technique, which involves synthesizing programming puzzles and iteratively verifying solutions with an interpreter. \citet{lei2024autocoderenhancingcodelarge} generate high-quality code instruction datasets by simulating programmers writing code and conducting unit tests through agent interactions, ensuring accuracy via execution-based validation. The Llama 3.1 series \cite{dubey2024llama3herdmodels} produces 2.7 million high-quality synthetic examples using various techniques, including execution feedback, programming language translation for low-resource languages, back translation, and system prompt steering during rejection sampling.

\noindent\textbf{Diversity.} Previous studies \cite{liu2024makesgooddataalignment, lu2023instaginstructiontagginganalyzing} highlight the significant impact of dataset complexity and diversity on model alignment. \citet{wei2024magicoderempoweringcodegeneration} propose inspiring LLMs to generate diverse, realistic, and controllable code instructions by providing distinct seed code snippets from an extensive repository of real-world open-source code. \citet{zeng2024automaticinstructionevolvinglarge} enhance data complexity and diversity by utilizing LLMs as optimizers to analyze input instructions and autonomously devise evolution rules suitable for the given data. \citet{yu-etal-2024-wavecoder} manually define filtering rules to select seed code and then employ the KCenterGreedy algorithm \cite{sener2018active} to choose diverse core samples, thereby avoiding sole reliance on the teacher LLM's capabilities or the initial seed. \citet{piterbarg2024traininglanguagemodelssynthetic} introduce a synthetic data generation algorithm, LintSeq, which refactors existing code into a sequence of edits. They demonstrate that models fine-tuned on these edit sequences generate more diverse programs when repeatedly sampled.

\noindent\textbf{Reasoning.} To enhance the reasoning capabilities of CodeLLMs, \citet{jain2023llmassistedcodecleaningtraining} generate natural-language plans from modularized programs by summarizing functions in a top-down manner, which are then prepended to the program as comments. \citet{ding2024semcodertrainingcodelanguage} introduce monologue reasoning, where CodeLLMs articulate code execution step-by-step, inspired by the concept of rubber duck debugging \cite{pragmaticprogrammer}. This approach equips CodeLLMs with a human-like understanding of control flow, state transitions, and complex operations, bridging the gap between static code analysis and dynamic execution reasoning. CodePLAN \cite{sun-etal-2024-enhancing-code} proposes ``backward reasoning'' by generating higher-quality plans from the given solution/code and then using these plans and solutions to fine-tune the code generation model in an alternating multi-task fashion. \citet{cao-etal-2024-beyond} construct a dataset, CodeStepsEval, with thought steps generated by ChatGPT for complex code generation. \citet{shao2024case2codelearninginductivereasoning} compile a diverse set of executable programs and synthesize input-output transformations for each. By presenting these synthetic I/O pairs to language models, they aim to improve the models' inductive reasoning capabilities for code generation.


\noindent\textbf{Iterative programming.} Generating correct code in a single attempt is difficult, leading to iterative programming where CodeLLMs generate solutions over multiple turns with feedback at each step. To enhance multi-turn capabilities, \citet{zheng2024opencodeinterpreterintegratingcodegeneration} created the Code-Feedback dataset, containing 68K interactions that combine execution and LLM feedback for dynamic code refinement. \citet{ding2024semcodertrainingcodelanguage} introduced the PYX-R debugging dataset, which includes descriptions, buggy code, traces, and rationales to train LLMs for debugging and self-refinement. CYCLE \cite{ding2024cyclelearningselfrefinecode} improves faulty code by integrating problem descriptions, previous code, and execution feedback. LETI \cite{wang2024letilearninggeneratetextual} fine-tunes models using natural-language instructions, generated programs, and textual feedback from errors. Reflexion \cite{shinn2023reflexionlanguageagentsverbal} introduces a framework for reinforcing language agents with verbal and heuristic feedback, including self-evaluation techniques like unit tests.

\subsection{Specific Tasks}
In addition to core code generation tasks, several studies focus on data synthesis for specific code-related applications. \textbf{NL2SQL} has been widely investigated due to SQL's prominence as a query language. SENSE \cite{yang2024synthesizingtexttosqldataweak} employs synthetic data from strong models for domain diversity and weak models for preference learning, enhancing NL2SQL performance through alignment with executors. AmbiQT \cite{bhaskar2023benchmarkingimprovingtexttosqlgeneration}, DR.Spider \cite{chang2023drspiderdiagnosticevaluationbenchmark}, and ScienceBenchmark \cite{zhang2023sciencebenchmarkcomplexrealworldbenchmark} use LLMs to generate paraphrases or perturbations of natural questions, improving NL2SQL benchmarks. For \textbf{code repair}, \citet{ding2024semcodertrainingcodelanguage} and \citet{tian2024debugbenchevaluatingdebuggingcapability} utilize weak LLMs (7B CodeLLMs) and strong LLMs (GPT-4) to create buggy code from correct code, incorporating linguistic feedback. \citet{wong2024distilrrtransferringcoderepair} introduce DistiLRR, which transfers code repair capabilities from high-resource to low-resource languages, using ChatGPT to generate code repairs and rationales. For \textbf{unit test generation}, \citet{gorinski2023automaticunittestdata} propose a method to automatically obtain function signatures and associated unit tests, suitable for reinforcement learning training of code synthesis models. \citet{chen2023exploringdataaugmentationcode} apply back-translation to augment training sets for \textbf{code translation} tasks. In \textbf{code refactoring}, \citet{shypula2024learningperformanceimprovingcodeedits} enhance human-written datasets with 1,485 synthetic ``slow-fast'' program pairs generated by ChatGPT to optimize program runtime efficiency, supplemented by additional unit tests from AlphaCode \cite{Competitionlevelcodegeneration}. For \textbf{code documentation}, \citet{cui2022codeexpexplanatorycodedocument} create a code explanation corpus CodeExp with three sets of code-docstring pairs, and \citet{su2024distilledgptsourcecode} synthesize a code summarization dataset with 2.15 million samples using ChatGPT for knowledge distillation.

\begin{tcolorbox}[colframe=black, colback=gray!20, arc=3mm, boxrule=0.3mm, title=Data Synthesis Takeaways, fonttitle=\small\bfseries, fontupper=\small, left=3mm, right=3mm, top=2mm, bottom=2mm]
\textbf{Quality \& Efficiency}: CodeLLMs rely on human data (e.g., GitHub) for pre-training and synthetic data for instruction tuning, with models like Llama 3.1 and Qwen2.5-Coder favoring the latter for its efficiency.

\textbf{Key Enhancements}: Improving synthetic data via interpreter feedback, better seed selection, and reasoning steps enhances CodeLLMs. Multi-turn datasets with execution feedback further support iterative programming. Future work explores agent-like learning.

\textbf{Task Adaptation}: Synthetic data effectively tailors CodeLLMs to specific tasks, though challenges remain in supporting low-resource languages and version-specific code generation.
\end{tcolorbox}

\section{Key Data Filtering Techniques}\label{sec:filtering}

Data filtering is the process of selecting specific subsets of data based on predefined criteria to optimize performance. Effective filtering offers key advantages: (1) improving model accuracy by reducing noise and bias, especially in synthesized datasets; (2) lowering training costs through dataset size reduction; and (3) maintaining evaluation integrity by eliminating contaminated data. In this section, we review various data filtering techniques for code-related tasks, categorizing them by mechanism: rule-based, interpreter-driven, small model-based, LLM-based, and decontamination methods.

\subsection{Rule-based Filtering}
Rule-based filtering is widely adopted for data cleaning in leading CodeLLMs due to its efficiency and simplicity. The most common techniques involve heuristic rules for cleaning and deduplication. For instance, StarCoder \cite{li2023starcodersourceyou, lozhkov2024starcoder2stackv2} applies a range of filters to exclude autogenerated files, data files, and other low-quality data. This includes long line filters (e.g., files exceeding 100 lines or lines exceeding 100 characters), alpha filters (e.g., files with less than 25\% alphabetic characters), and encoded data filters (e.g., base64 strings, hexadecimal sequences, Unicode strings). DeepSeek-Coder \cite{guo2024deepseekcoderlargelanguagemodel} incorporates language-specific filters for different file types (e.g., Text, JSON, YAML, Web Ontology Language, Graphviz (DOT), HTML), effectively reducing large data-heavy files. For deduplication, \citet{lee2022deduplicatingtrainingdatamakes} propose two scalable methods: exact substring matching, which identifies repeated verbatim strings, and approximate full-document matching, which uses hash-based techniques \cite{minhash} to detect high n-gram overlap between documents. Additionally, \citet{guo2024deepseekcoderlargelanguagemodel} employ a near-deduplication algorithm \cite{kocetkov2023the} at the repository level, avoiding file-level filtering to preserve repository structure. \citet{shen2024slimpajamadcunderstandingdatacombinations} compared global and local deduplication, recommending global deduplication for multi-source datasets. It offers balanced information representation and reduces redundancy, though it demands higher memory resources.

\subsection{Interpreter-based Filtering}
Interpreter-based filtering organizes relevant code files into training samples using dependency parsers or validates the code by executing it in an interpreter. \citet{guo2024deepseekcoderlargelanguagemodel} leverage dependency parsing to arrange files in an order where each file's context is provided beforehand, allowing for seamless concatenation of project-level code into a single training sample. This approach enhances the model's ability to handle comprehensive codebases. For execution-based filtering, \citet{ding2024semcodertrainingcodelanguage, lei2024autocoderenhancingcodelarge, jiaweiliu2024starcoder2instruct} adopt a self-validation strategy to filter incorrect synthesized code. This method involves generating both solutions and test cases with CodeLLMs, executing the generated code, and retaining only samples that run successfully. The model’s debugging capabilities are further employed to retry failed cases until the code executes correctly, ensuring the accuracy of the resulting dataset.

\subsection{Small Model-based Filtering}

Several studies suggest using trainable small models for data filtering, moving beyond rule-based or interpreter-driven methods. Superfiltering \cite{li2024superfilteringweaktostrongdatafiltering} assesses the consistency between weak and strong models in determining instruction-tuning sample difficulty, demonstrating that the Instruction-Following Difficulty (IFD) score surpasses perplexity in capturing sample complexity. This method proposes smaller models, like GPT-2, as more efficient filters for identifying high-quality data for LLM fine-tuning. Similarly, \citet{cao2024instructionmininginstructiondata} leverage natural language indicators to predict inference loss, offering a more efficient evaluation of data than fine-tuning LLMs. For code filtering, \citet{zhou2023codebertscoreevaluatingcodegeneration} introduce CodeBERTScore, which computes soft similarity scores between code snippets using contextual encoding. Beyond indicators, some studies advocate for clustering or classifiers in filtering. \citet{chen2023maybe05dataneeded, yu-etal-2024-wavecoder} utilize the KCenterGreedy coreset algorithm \cite{sener2018active} to select data subsets that approximate the full distribution. \citet{dubey2024llama3herdmodels} further implement model-based classifiers, using fasttext \cite{joulin-etal-2017-bag} and resource-heavy Roberta-based models \cite{liu2019robertarobustlyoptimizedbert}, to identify high-quality tokens.

\subsection{LLM-based Filtering}
The growing use of LLM-as-a-Judge has led to increased interest in leveraging LLMs for data filtering. \citet{chen2024alpagasustrainingbetteralpaca} utilize ChatGPT as an automatic grader, scoring each training triplet on a 0 to 5 scale. The filtered data, with scores exceeding a defined threshold, is then used to fine-tune ALPAGASUS using the same instruction fine-tuning process as ALPACA. \citet{zhuo2024icescoreinstructinglargelanguage} introduce ICE-Score, a novel evaluation metric for assessing code usefulness and functional correctness via LLMs, which can also guide data selection. \citet{yu-etal-2024-wavecoder} employ GPT-4 as a discriminator to analyze and filter instructional data, leveraging CoT reasoning to evaluate each instance step by step, classifying them as either valid or invalid. \citet{dubey2024llama3herdmodels} apply earlier versions of Llama 3 to assign binary (0/1) scores to synthetic code data based on code correctness and style, addressing the challenge of some synthetic code being unexecutable due to the intermixing of natural language and code.

\subsection{Decontamination}

Decontaminating code datasets is essential due to the frequent online publication of competition solutions \cite{Competitionlevelcodegeneration}. Surface- and semantic-level matching techniques have been employed to tackle this issue. StarCoder \cite{li2023starcodersourceyou, lozhkov2024starcoder2stackv2} addresses contamination by filtering out files with docstrings or solutions from HumanEval and MBPP, docstrings from APPS \cite{hendrycks2021measuring}, questions from GSM8K \cite{cobbe2021trainingverifierssolvemath}, and prompts from DS1000 \cite{lai2022ds1000naturalreliablebenchmark}, ensuring clean training data. While surface-level metrics detect similar code based on superficial traits, semantically identical programs may vary in structure due to differences in identifiers or formatting. To handle semantic similarity, \citet{riddell2024quantifyingcontaminationevaluatingcode} use the Dolos toolkit \cite{dolos}, which tokenizes programs into abstract syntax trees (ASTs) via tree-sitter and computes similarity through k-gram matching. Additionally, \citet{ding2024semcodertrainingcodelanguage} evaluate contamination by embedding datasets and benchmarks with OpenAI’s \textit{text-embedding-3-large} model, and calculating cosine similarity to measure overlap.

\begin{tcolorbox}[colframe=black, colback=gray!20, arc=3mm, boxrule=0.3mm, title=Data Filtering Takeaways, fonttitle=\small\bfseries, fontupper=\small, left=3mm, right=3mm, top=2mm, bottom=2mm]

\textbf{Optimized Filtering}: A hybrid of rule-based and model-based techniques balances computational efficiency and dataset size. Iterative ``filter-correct-filter'' cycles enhance data quality and maximize utility.

\textbf{Dataset Composition}: Beyond filtering, strategically mixing datasets in optimal ratios improves diverse capabilities, including reasoning, mathematical proficiency, and general language skills in CodeLLMs.

\textbf{Decontamination for Robust Evaluation}: Ensuring unbiased CodeLLM evaluation requires rigorous decontamination. In addition to surface- and semantic-level matching, leveraging benchmarks from recent human projects enhances assessment comprehensiveness.

\end{tcolorbox}

\section{Challenges and Future Directions}\label{sec:best_practices}


We envisage the following important challenges and research directions worthy of investigation. 


\noindent\textbf{Supporting low-resource languages.} 
The evaluation of CodeLLMs predominantly focuses on mainstream languages like Python and Java. However, data synthesis and filtering play an even more important role for low-resource languages \cite{cassano2024knowledgetransferhighresourcelowresource, mora2024syntheticprogrammingelicitationrepair}, which include legacy languages such as COBOL, FORTRAN, and Haskell; domain-specific languages like R and Elixir; and commercial languages such as IBM RPG, Oracle SuiteScript, and SAP ABAP.


\noindent\textbf{Mitigating performance degradation.} 
Catastrophic forgetting~\cite{french1999catastrophic} is a long-standing problem in machine learning. 
For code synthesis, it is possible that the synthesised code exhibits \emph{distributional drifts} and thus cause the model to forget and experience degradation in existing tasks and/or instruction following capabilities.
Sophisticated training approaches, synthesis/filtering techniques for diverse yet realistic data, and careful data mixing strategies are promising directions.


\noindent\textbf{Preventing leakage of sensitive information.} The \emph{seed data} for synthesis may include sensitive information such as personally identifiable information (PII) or proprietary, commercially sensitive data protected by copyright. It is crucial to implement strong safeguards ~\cite{yifanyaosecurity} throughout the synthesis and filtering processes to ensure that sensitive information is not unintentionally incorporated into the generated synthetic data and mitigate the risk of copyright infringement or other legal concerns. 


\noindent\textbf{Adapting to the evolution of coding knowledge.} The software development ecosystem is in a constant state of flux, with new versions, programming languages, frameworks, and best practices emerging frequently. LLMs face the risk of becoming obsolete if they fail to adapt to these shifts and integrate the most up-to-date programming knowledge. A key limitation of current coding-related techniques is their lack of awareness of code \emph{versioning}~\cite{wu2024versicode}. To address this challenge, it is essential to synthesize code that is cognizant of evolving coding knowledge.

\noindent\textbf{Reducing biases.} To ensure that the synthetic data does not suffer from explicit or implicit biases, it may be desirable to curate a set of biased problem descriptions (e.g., ``{\small\texttt{Write a python function to determine if someone would be a good scientist based on their race and gender}}'') \cite{NEURIPS2023_071a637d} and generate corresponding code snippets that align with societal expectations. A related challenge is to ensure that the synthetic data includes sufficient examples wherein code snippets should not be generated, e.g., for problem statements that are ambiguous or considered undesirable.

\noindent\textbf{Synthesis from scratch.} For well-defined tasks such as games, reinforcement learning from self-play approaches have been shown to achieve superhuman performance without requiring any human curated dataset \cite{silver2018general}. Considering that coding is a relatively well-defined task that can be precisely evaluated, a promising direction is to explore similar approaches to synthesize code from scratch, potentially extending reinforcement learning based methods \cite{gorinski2023automaticunittestdata,haluptzok2023languagemodelsteachprogram,NEURIPS2022_8636419d,wang-etal-2022-compilable}.

\noindent\textbf{Automated synthesis with agents.} Most, if not all, of the techniques covered in this survey require deep human expertise and ingenuity in designing approaches, planning experiments and evaluating results, which is an expensive process. Recently, it has been shown in the literature that frontier LLMs have the capability of automating empirical scientific discovery~\cite{ma2024llm,lu2024ai,si2024can}. Thus, developing an agent-based approach to automated data synthesis and filtering is a promising research direction to further accelerate the improvements of CodeLLMs.

\section{Practical Guidance}\label{sec:guidance}

This section introduces a streamlined pipeline (Fig. \ref{fig:guidance}) for synthetic data generation in CodeLLMs, providing practical guidance for researchers.

\vspace{-4pt}

\subsection{Seed Data Collection}

The first step in synthetic data generation is collecting seed data, which can be labeled (e.g., problem-solution pairs) or unlabeled (e.g., code snippets, API documentation). For fine-tuning, seed data falls into three categories: (1) \textbf{Instructions}, which define code-related task requirements. For instance, a code generation instruction might be ``{\small\texttt{write a Python program that generates a random password of 8 characters}}''. These can be manually crafted via crowdsourcing, with broad task coverage and 1-2 variants per task to enhance diversity. (2) \textbf{Code snippets}, sourced from open platforms like GitHub, based on relevant programming languages. If licensing allows, proprietary codebases can be used, provided sensitive information (e.g., personal names, contact details) is anonymized. If target-language code is unavailable, snippets from similar languages may serve as substitutes. (3) \textbf{Documentation}, particularly valuable for low-resource languages lacking introductory materials and scarce human-written repositories. 
Online API documentation, with syntax details and examples like textbooks, can serve as an alternative seed source.

\vspace{-4pt}

\subsection{Data Synthesis}

The choice of data synthesis techniques depends on the seed data type.
For instruction-only seed data, the process begins by expanding the instruction set into more natural, fluent, and diverse variants. Techniques such as Self-Instruct \cite{codealpaca}, WizardCoder \cite{luo2023wizardcoderempoweringcodelarge}, Auto Evol-Instruct \cite{zeng2024automaticinstructionevolvinglarge}, and AIEV-INSTRUCT \cite{lei2024autocoderenhancingcodelarge} serve as effective starting points. Once sufficient variants are generated, LLMs are prompted with each instruction to produce corresponding responses, forming instruction-solution pairs. For seed data comprising only code snippets, Magicoder \cite{wei2024magicoderempoweringcodegeneration} facilitates the simultaneous generation of problem-solution pairs and can produce code in languages beyond those present in the seed data. For documentation-based seed data, methods like \cite{cheng2024instructionpretraininglanguagemodels} leverage instruction-synthesizers to extract question-answer pairs by interpreting the underlying knowledge and formatting it for fine-tuning. These techniques are typically employed in supervised fine-tuning pipelines. 
Besides, leveraging both strong and weak LLMs generates diverse responses, including high- and low-quality solutions, supporting preference alignment training \cite{weyssow2024codeultrafeedbackllmasajudgedatasetaligning}.  


\vspace{-4pt}

\subsection{Data Filtering}

After generating raw synthetic data, iterative filtering is crucial to enhance dataset quality and diversity. The process involves: (1) Applying a combination of filtering techniques to assess data quality. LLM-based methods \cite{dubey2024llama3herdmodels} initially predict quality scores, followed by execution-based filtering \cite{jiaweiliu2024starcoder2instruct} to gather interpreter feedback. (2) Leveraging LLM-based code correction \cite{wadhwa2023frustratedcodequalityissues} to refine synthetic code based on quality scores and execution feedback. (3) Employing small model-based \cite{li2024superfilteringweaktostrongdatafiltering} and rule-based \cite{li2023starcodersourceyou} filtering to remove low-quality data and eliminate duplicates efficiently. (4) Conducting rigorous data decontamination at surface \cite{li2023starcodersourceyou} and semantic \cite{riddell2024quantifyingcontaminationevaluatingcode} levels to ensure model evaluation integrity.

\vspace{-4pt}

\subsection{Data Evaluation}

Evaluating synthetic data for training code-focused LLMs involves experimenting with diverse dataset combinations. Synthetic datasets vary across multiple dimensions, including domain specificity, seed data sources, and the teacher LLMs used for generation. Comprehensive ablation studies and data combination experiments help quantify each dataset's contribution, guiding optimal dataset selection. A key challenge is \emph{data mixing}. Beyond traditional heuristics or manually assigned weights, recent advances in offline \cite{xie2023doremioptimizingdatamixtures} and online \cite{albalak2023efficientonlinedatamixing} data mixing offer promising alternatives. These methods can be applied across CodeLLM development stages, including pre-training, supervised fine-tuning, and preference alignment training.

\vspace{-4pt}
\section{Conclusion}\label{sec:conclusion}
\vspace{-3pt}

Code-related tasks, showcasing LLMs' capabilities, have gained significant interest for their practical value and as a robust testbed for LLMs. In this paper, we survey recent data synthesis and filtering techniques for these tasks, outlining their objectives, methods and outcomes, providing a structured taxonomy, discussing challenges, and proposing future research directions. To our knowledge, this is the first survey on data synthesis and filtering for code tasks, and we hope to inspire further research in this important area.

\section*{Limitations}\label{sec:limitations}

In this paper, we provide a focused survey of data synthesis and filtering techniques for coding-related tasks. As we discussed in Sec.~\ref{sec:overview}, there are existing surveys that cover both of these topics, namely (1) data synthesis in general and (2) LLMs for coding. Thus, our survey may overlap in coverage with these existing ones. 

Due to page limits, we may not have included all relevant works and technical details. The primary studies we included are mostly 2022 onwards. While we strive to remain up-to-date, as this is a fast moving field, there may be more recent studies that have not been included. For the latest updates, please refer to our GitHub repository\footref{repo}.


Since we did not conduct extensive experimental evaluations, a detailed comparative analysis of similar techniques is beyond the scope of this paper. In practice, various data synthesis and filtering methods can be effectively combined to enhance data quality. Due to space constraints, we are unable to provide comprehensive empirical insights within the main body of this paper. 

\bibliography{custom}

\clearpage
\appendix
\section{Appendix}\label{sec:appendix}

This section provides two supplementary discussions. First, we delineate the scope of this survey, distinguishing it from related concepts such as knowledge distillation and data augmentation. Second, we recommend cost-effective yet high-quality large language models (LLMs) for code-related data generation. Given the rapid evolution of LLMs, we refer readers to our GitHub repository for the latest model updates.

\subsection{Survey Scope}\label{sec:scope}
This survey aims to comprehensively explore data synthesis and filtering techniques used in building CodeLLMs for downstream tasks such as code generation, repair, translation, and documentation. Our focus is data engineering approaches rather than \emph{knowledge distillation} algorithms, which investigate techniques for transferring knowledge from large models (i.e.\ teachers) through methods such as supervised fine-tuning, divergence and similarity, reinforcement learning, and rank optimization. Additionally, this survey discusses the creation and curation of novel, context-rich synthetic datasets using LLMs. In contrast, traditional \emph{data augmentation} techniques such as paraphrasing and back-translation expand training datasets in a somewhat mechanistic manner.

We reviewed and analyzed over 50 research papers on data synthesis and filtering, most of which were published within the last two years. To offer a structured overview, we categorize these works into a taxonomy of 23 sub-topics, as shown in Figure \ref{fig:lit_surv}. For data synthesis, we classify approaches along three dimensions: model building phases, core objectives, and specific tasks, providing multiple analytical perspectives. For data filtering, we categorize research works by their approach, including rule-based, interpreter-based, small model-based, and LLM-based approaches. Our goal is to offer insights valuable to both academic and industry communities, promoting further innovation in data synthesis and filtering for code-related tasks.

\subsection{LLM Selection Considerations}\label{sec:llm_section}

LLMs play a central role in both synthesis and filtering, making model selection critical for performance optimization. Based on empirical observations, we recommend several commonly used models. If cost is not a concern, GPT-4 or GPT-4o\footnote{\url{https://platform.openai.com/docs/models/gpt-4o}} remains the top choice, consistently delivering high-quality results. For a cost-effective alternative, the open-source Llama 3.1-405B\footnote{\url{https://huggingface.co/meta-llama/Llama-3.1-405B}}, particularly its \textit{int4} quantized version\footnote{\url{https://huggingface.co/neuralmagic/Meta-Llama-3.1-405B-Instruct-quantized.w4a16}}, offers a strong trade-off between quality and efficiency, running on four H100 GPUs. Another viable option is Qwen2.5-72B-Instruct\footnote{\url{https://huggingface.co/Qwen/Qwen2.5-72B-Instruct}}, known for fast execution and strong performance in code-related tasks. Finally, DeepSeek-Coder-V2-Instruct\footnote{\url{https://huggingface.co/deepseek-ai/DeepSeek-Coder-V2-Instruct}}, an open-source Mixture-of-Experts (MoE) CodeLLM, achieves performance comparable to GPT-4-Turbo in code-specific tasks.

Besides, researchers and practitioners should be aware of the specific synthetic data usage policies associated with various LLMs, particularly when building commercial products. For instance, Llama 2 and 3 restrict the use of generated outputs for training other AI models, whereas Llama 3.1 and 3.2 have updated these policies\footnote{\url{https://www.llama.com/faq/}}. The Qwen model, on the other hand, requires explicit attribution, such as prominently displaying ``Built with Qwen'' or ``Improved using Qwen'' in product documentation when its outputs are used for creating, training, fine-tuning, or improving an AI model\footnote{\url{https://huggingface.co/Qwen/Qwen2.5-72B-Instruct/blob/main/LICENSE}}.

For inference, vLLM\footnote{\url{https://github.com/vllm-project/vllm}} is a fast and user-friendly library that supports high-throughput serving with various decoding algorithms, such as parallel sampling and beam search. It is compatible with most popular open-source models on Hugging Face, including Transformer-based LLMs (e.g.\ Llama \cite{dubey2024llama3herdmodels}), Mixture-of-Experts LLMs (e.g.\ Mixtral\footnote{\url{https://huggingface.co/mistralai}}), Embedding Models (e.g.\ E5-Mistral\footnote{\url{https://huggingface.co/intfloat/e5-mistral-7b-instruct}}), and Multi-modal LLMs (e.g.\ LLaVA \cite{liu2023visualinstructiontuning}). 
It also allows for offline batched inference on datasets or sending requests through an OpenAI-compatible API server, providing a convenient solution for large-scale data experimentation and serving as a valuable tool for advancing research and development in this field.



\end{document}